\documentclass[aps,prl,preprint,groupedaddress]{revtex4}
\usepackage{amsmath,amssymb}
\usepackage{amsbsy}
\usepackage[utf8]{inputenc}
\usepackage[english]{babel}
\usepackage{graphicx}
\usepackage{subfigure}

\usepackage[a4paper,bindingoffset=0.2in,%
            left=.2 in,right=.2 in,top=1in,bottom=1in,%
            footskip=.25in]{geometry}

\def\bea{\begin{eqnarray}}
\def\eea{\end{eqnarray}}
\def\be{\begin{equation}}
\def\ee{\end{equation}}
\def\nn{\nonumber}

\def\m{\mu}
\def\n{\nu}

\def\om{\omega}

\def\a{\alpha}
\def\b{\beta}

\def\p{\partial}
\begin{document}

\title{ Interaction of Electromagnetic field with a Gravitational wave in Minkowski and de-Sitter space-time. }
\author{Akash Patel, }
\author{Arundhati Dasgupta
}

\affiliation{Physics and Astronomy, University of Lethbridge, 4401 University Drive, Lethbridge, Canada T1K 3M4.}
\email{arundhati.dasgupta@uleth.ca}

\email{a.patel2@uleth.ca}
\author{}
\affiliation{}

\begin{abstract}

It is important to re-examine some of the interactions of gravitational waves with matter due to the recent discovery of the waves in LIGO.  In a previous paper with Morales, interaction of gravitational waves was studied with scalar and neutrino fields in Minkowski space-time. In this paper, we find that Electromagnetic waves have a similar interaction with gravitational wave in a flat background, and we comment on the implications of this for a gravitational wave detector. Primordial gravitational waves were generated in the early universe and thus it is important to study interaction of the gravitational waves in cosmological backgrounds. In this paper we discuss the interaction of electromagnetic waves and gravitational waves in a de-Sitter background. We find that the Electromagnetic wave is perturbed and modulated. This result is new and will be useful in studying inhomogeneities in Cosmic Microwave Background. 	
 \end{abstract}


\maketitle

\section{Introduction}
In the recent years gravitational wave detection with LIGO \cite{ligo} has opened a new window for the study of the universe. Gravitational waves from distant events are arriving on earth and are being studied and analyzed using the LIGO detector. The Space Laser Interferometer will also be functional in the near future. In this circumstance, it is important to study the interaction of gravitational waves with matter, and predict new observations. Further, primordial gravitational waves have not been detected yet and new interactions might shed some light on the stochastic relics from early universe cosmology. For recent observations on the background gravitational waves see the NanoGrav experiment \cite{nano}.  In this paper we analyze and confirm a  perturbation for electromagnetic (EM) waves induced by gravitational waves, previously observed for scalars and neutrinos in \cite{morales} and in \cite{tsagas,chou}. We discuss resonant interactions, and if this new `frequency' mode as predicted can be detected in current detectors. Although our results are similar to \cite{tsagas,chou}, the analysis of resonance is different. We find no interactions when the gravitational waves and the EM waves are propagating in the same direction. When the waves are anti-parallel or at an angle, solutions are obtained for the perturbation, and we discuss the resultant EM wave. We find that the initial plane wave of the EM radiation changes direction and flows at an angle to the initial wave. The shape of the wavefront is also perturbed. However, the perturbation is very weak compared to the initial wave. The perturbation shows the `phase shift'  which can be identified with that causing the interference pattern observed in LIGO. In addition we predict a new mode, which can be observed if the frequency of the EM wave is of the order of the frequency of the gravitational wave. In the current detectors, the frequency of the EM laser is about $10^{14}$ Hz and the frequency of the gravitational wave $\sim 1-100$ Hz. Therefore, the EM frequency dominates in the observations. The work on designing an experiment to detect the mode is in progress with Maher \cite{maher}. 

In this paper we also discuss the interaction of gravitational waves with EM waves in a conformally flat cosmological background. We solve for the Maxwell's equation in the background of conformal de-Sitter space-time. We take this particular cosmology as this is a candidate space-time for early universe inflationary scenario, and also for maximally symmetric space-times with a positive cosmological constant.  We find that the EM perturbations grow quadratically with conformal time, and change in direction. We comment on the plausible detection of this perturbation as a inhomogeneity in the current cosmic microwave background.

In the next section we discuss the interaction of the EM waves with gravitational waves in a flat space-time. We find the perturbations and analyze the Poynting vector. We discuss the interactions when the waves are perpendicular to each other, and when they are in the same or opposite direction. In the third section we discuss the interaction of the EM waves with a gravitational wave in de-Sitter space background. We show that the resultant perturbations grow quadratically in conformal time whereas the zeroeth order gauge fields grow linearly in time. The fourth section is a conclusion.  In this paper we work in the units of Newton's constant $G$ and speed of light $c$, $G=c=1$. 

\section{Interaction of Electromagnetic Waves with Gravitational Waves in Minkowski space-time}
In this section we study the interaction of a gravitational wave travelling in the z-direction, and an electromagnetic wave in the x-direction. Due to the perpendicular nature of the direction of propagation, in linear analysis these waves should propagate independently of each other. However, due to the nature of the interactions of the electromagnetic wave and the gravitational wave, we find using a perturbation analysis that the electromagnetic wave gets non-trivially changed. We refrain from studying the back reaction of the electromagnetic wave on the gravitational wave, but that can also be found. At this time, we simply find the perturbation of the EM wave and predict how this perturbation might be detected at the LIGO detector. Initially we assume that there is a gravitational wave $h_{\mu \nu}$ with polarizations $h_{+}, h_{\times}$ propagating at a frequency $\omega_g$ in the z-direction. The amplitudes are taken as $A_{+}$ and $A_{\times}$. The gravitational wave metric is taken as \cite{liv}
\be
\left(\begin{array}{cccc}-1&0&0&0\\0&1+A_+\cos(\omega_g(z-t))&A_{\times}\cos(\omega_g(z-t))&0\\0&A_{\times}\cos(\omega_g(z-t))&1-A_+\cos(\omega_g(z-t))&0\\ 0&0&0&1\end{array}\right)
\ee
The electromagnetic wave is taken to be in the x -direction such that the magnetic field is 
\be
 \vec{B} = B_{0y}\ Re\{e^{i \omega_e( x- t)} \   \}\hat{y}
 \ee
 where $B_{0y}$ is its amplitude, and $\omega_e$ is its frequency.
  In the Lorenz Gauge the gauge field is found to be
 \be
 \vec{A}=-\frac{1}\omega_e B_{0y} \sin(\omega_e x-\omega_e t) \ \hat{z}.
 \ee  
 The field strength is ($\mu,\nu=0,1,2,3$)
 \be
 F_{\mu \nu}= \p_{\mu} A_{\nu} - \p_{\nu} A_{\mu}.
 \ee
 The non-zero components of the EM tensor is $F_{03}$ and (we use the notation t,x,y,z and 0,1,2,3 to represent the same coordinate directions interchangeably according to the convenience of the equations.) 
 \be
   F_{13}= -B_{0y} \cos(\omega_e(x-t)).
 \ee

 

 We solve the Maxwell's equation in the background of the gravitational wave which is as following:
 \be
 \frac{1}{\sqrt{-g}} \p_{\mu} \left(\sqrt{-g} g^{\mu \alpha} g^{\nu \beta} F_{\alpha \beta}\right)=0.
 \ee
 Using $g^{\mu \nu}= \eta^{\mu \nu} - h^{\mu \nu}$ , the inverted metric, we get from the above
{\small  \be
 \eta^{\mu \alpha} \eta^{\nu \beta} \p_{\mu} F_{\alpha \beta} - \p_{\mu} (h^{\mu \alpha}) \eta^{\nu \b} F_{\a \b} -\p_{\mu} (h^{\nu \a})\eta^{\mu \b}F_{\a \b} - h^{\m \a}\eta^{\n \b} \p_{\m} F_{\a \b} - \eta^{\m \a}h^{\n \b}\p_{\mu} F_{\a \b}=0.
\label{eqn:maxwell} \ee}
 We then assume a perturbation of the initial gauge field $A_{\mu}= \bar{A}_{\mu}+ \tilde{A}_{\mu}$ (where $\bar{A}$ represents the wave without the gravitational wave and $\tilde{A}$ is the perturbation) and keep terms to linear order in $\tilde{A}$ and $h$ in Equation(\ref{eqn:maxwell}). Using the Lorenz gauge conditions $\p_{\mu} A^{\mu}=0$ and $\p^{\mu}h_{\mu \alpha}=0$;  the perturbation equations are
 \bea
 \Box \tilde{A}_0 & = & 0\\
 \Box \tilde{A}_1 & =  &(\p_3 h^{11}) F_{31} = - A_{+} B_{0y} \  \omega_g \sin(\omega_g z- \omega_g t) \cos(\omega_e x- \omega_e t) \\ 
 \Box \tilde{A}_2&=& \p_3(h^{21} F_{13}) = A_{\times} B_{0y} \ \omega_g \sin(\omega_g z- \om_g t + \delta) \cos(\om_e x-\om_e t)\\
 \Box \tilde{A}_3 & = & h^{11} \p_1 F_{13} = A_{+} B_{0y} \ \om_e \sin(\om_e x- \om_e t) \cos(\om_g z - \om_g t) .\eea
 ($\square A_{\mu}$ is the relativistic d'Alembertian acting on the gauge field.)\\
 Note that the inhomogeneous differential equations in the linearized approximation is natural, as we are solving for Maxwell's equation in curved geometry. The curvature of the geometry provides the `driving force' for the perturbations. The inhomogeneous terms can be interpreted as a `current source' proportional to the $B_{0y} A_+$, product of the EM wave and the gravitational wave. This also signifies that if the initial EM wave is zero, there is no perturbation at all. The linearized approximation has been discussed in \cite{monta}, in analogy of a circuit, however, we differ from their analysis. In \cite{monta} it has been claimed that there has to be an `external source of emf' to justify the inhomogeneous terms in the above equations. However, circuits are known to have the capacity to generate their own emf, e.g.  `self inductance' . In a similar analogy, we can see that the there can be self generation of emf or `back emf' in the circuit due to the curvature of the background. Therefore linearized approximations which give source terms for the perturbations can be used to solve for interactions, and some of the answers obtained in \cite{monta} using Eikonal approximation is similar qualitatively to the results of this paper. It remains that experimental detection of the new results will confirm correctness of the approximations. 
 
 The perturbation solution is obtained using the methods of \cite{morales} and as outlined in the Appendix1. The x-component is found to be 
 \bea
 \tilde{A}_x &= &  -\frac{B_{0y} A_+ }{4 \omega_e \sqrt{\omega_g^2 + \omega_e^2}} \left[ (\omega_e + \omega_g)\cos(\omega_e x + \omega_g z) \sin (t \sqrt{\omega_e^2 + \omega_g^2}) - \sqrt{\omega_g^2 + \omega_e^2} \ \sin(\omega_e x + \omega_g z) \cos(t \sqrt{\omega_g^2 + \omega_e^2})\right. \nn \\
 &+ & (\omega_e -\omega_g) \cos(\omega_e x - \omega_g z) \sin (t \sqrt{\omega_e^2 + \omega_g^2}) - \sqrt{\omega_g^2 + \omega_e^2} \ \sin(\omega_e x -\omega_g z) \cos (t \sqrt{\omega_e^2 + \omega_g^2}) \nn \\
 &+& \left. \sqrt{\omega_e^2 +\omega_g^2} \ \sin(\omega_e x + \omega_g z - t(\omega_g + \omega_e) ) + \sqrt{\omega_e^2+\omega_g^2}\ \sin(\omega_e x - \omega_g z -t(\omega_e -\omega_g)) \right].
 \label{eqn:sol1}
 \eea
 The solution for $\tilde{A}_y$ would be the same as this, except for $-A_{+} \rightarrow A_{\times}$ and 
 $\omega_g z\rightarrow \omega_g z +\delta$ . The solution for $\tilde{A}_{z}$ will also be the similar; which we can obtain by using phase shifts of $\omega_e x\rightarrow  \omega_e x+ \pi/2$ and $\omega_g z \rightarrow \omega_g z + \pi/2$ and in the coefficient $- B_{0y} A_+ /(4 \omega_e \sqrt{\omega_e^2 + \omega_g^2}) \rightarrow B_{0y} A_+/ (4\omega_g\sqrt{\omega_e^2+\omega_g^2})$. Note that these solutions are obtained with the boundary condition that the perturbations are zero at time $t=0$. 
 
 It is interesting to see that there are terms of the form $\sin(\omega_e x\pm \omega_g z -t (\omega_g\pm \omega_e))$ which can be interpreted as causation of `phase shifts', and would lead to interference patterns, when incident on a detector with another EM wave. But there are also terms of the form $\cos(\sqrt{\omega_e^2+\omega_g^2}~t)$ which see a new frequency `mode' in the solution. These terms have been observed previously in \cite{tsagas,chou}.  Question is why haven't they been observed in the LIGO? The simple answer is that the LASER frequency used in the LIGO or $\omega_e \sim 10^{14}$ Hz, and the gravitational wave frequency is of the order of $1-100$ Hz. Therefore the frequency of the new mode $\omega\sim \omega_e$ and therefore one has to modify the details of the experiment to be able to see the mode. This is under investigation in \cite{maher}. Note that the
 above perturbations amplitude is inversely proportional to $\omega_e$ and thus bringing down the frequency of the EM wave in the interferometer will increase the strength of the signal.  The z-component of the gauge potential, which represents a EM wave perturbation over the original has an amplitude which is proportional to $1/\omega_g$ and therefore gives rise to a bigger amplitude wave than the other components. Methods to detect the perpendicular component of the Poynting vector is under investigation in \cite{maher}.
 The perturbed $\vec{E}$ and the $\vec{B}$ components of the fields from the above gauge potentials can be derived by computing 
 \be
 \vec{\tilde E}=- \frac{\partial \vec{\tilde A}}{\partial t}
 \ee
 and we can see that the perturbed field has components in all three dimensions. E.g. the x-component is (the y and z components can similarly be found)
 \bea
 \tilde{E}_x &= &\frac{B_{0y} A_+ }{4 \omega_e} \left[ (\omega_e + \omega_g)\cos(\omega_e x + \omega_g z) \cos(t \sqrt{\omega_e^2 + \omega_g^2}) + \sqrt{\omega_g^2 + \omega_e^2} \ \sin(\omega_e x + \omega_g z) \sin(t \sqrt{\omega_g^2 + \omega_e^2})\right. \nn \\
 &+ & (\omega_e -\omega_g) \cos(\omega_e x - \omega_g z) \cos (t \sqrt{\omega_e^2 + \omega_g^2}) + \sqrt{\omega_g^2 + \omega_e^2} \ \sin(\omega_e x -\omega_g z) \sin (t \sqrt{\omega_e^2 + \omega_g^2}) \nn \\
 &-& \left.  (\omega_g+\omega_e)\ \sin(\omega_e x + \omega_g z - t(\omega_g + \omega_e) ) - (\omega_e-\omega_g)\ \sin(\omega_e x - \omega_g z - t(\omega_e -\omega_g)) \right].
 \eea
 
 The Magnetic Field is:
 
 \be
 \vec{B}= \vec{\nabla}\times \vec{A}.
 \ee
 
 The newly generated z-component of the $\vec{B}$ field is hence given by
 \be
\tilde{ B}_z=\frac{\partial \tilde{A}_y}{\partial x}
 \ee
 \bea
\tilde{B}_z&=&   \frac{B_{0y} A_{\times} }{4  \sqrt{\omega_g^2 + \omega_e^2}} \left[ -(\omega_e + \omega_g)\sin(\omega_e x + \omega_g z+\delta) \sin (t \sqrt{\omega_e^2 + \omega_g^2}) \right. \nn \\ &-& \sqrt{\omega_g^2 + \omega_e^2} \ \cos(\omega_e x + \omega_g z+\delta) \cos(t \sqrt{\omega_g^2 + \omega_e^2}) \nn \\
 &- & (\omega_e -\omega_g) \sin(\omega_e x - \omega_g z+\delta) \sin (t \sqrt{\omega_e^2 + \omega_g^2}) - \sqrt{\omega_g^2 + \omega_e^2} \ \cos(\omega_e x -\omega_g z+\delta) \cos (t \sqrt{\omega_e^2 + \omega_g^2}) \nn \\
 &+& \left. \sqrt{\omega_e^2 +\omega_g^2} \ \cos(\omega_e x + \omega_g z+\delta - t(\omega_g + \omega_e) ) + \sqrt{\omega_e^2+\omega_g^2}\ \cos(\omega_e x - \omega_g z + \delta - t(\omega_e -\omega_g)) \right].
\eea
The x,y components can be similarly derived. 

\begin{figure}
\subfigure{\includegraphics[scale=0.2]{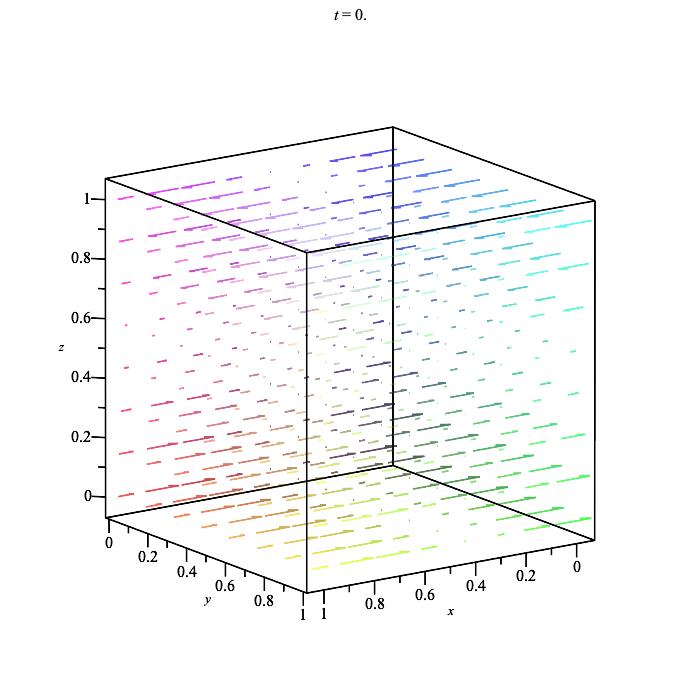}}
\subfigure{\includegraphics[scale=0.2]{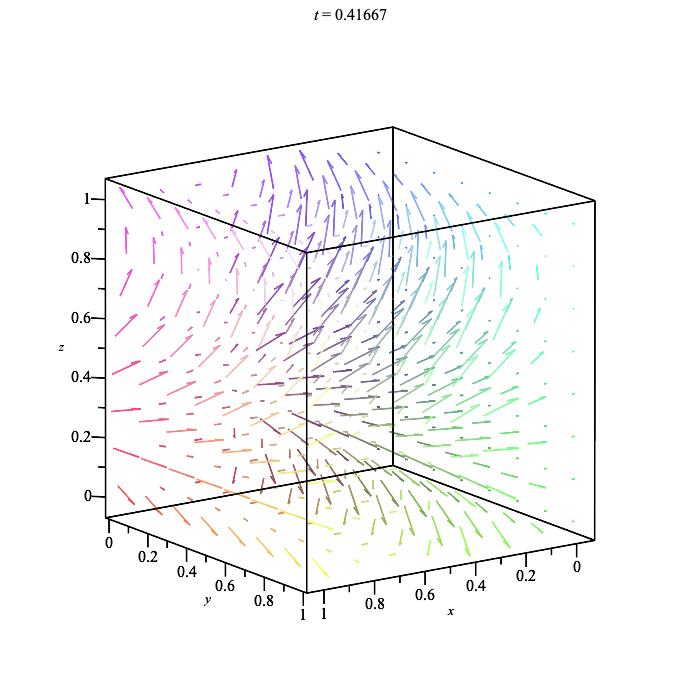}}
\subfigure{\includegraphics[scale=0.2]{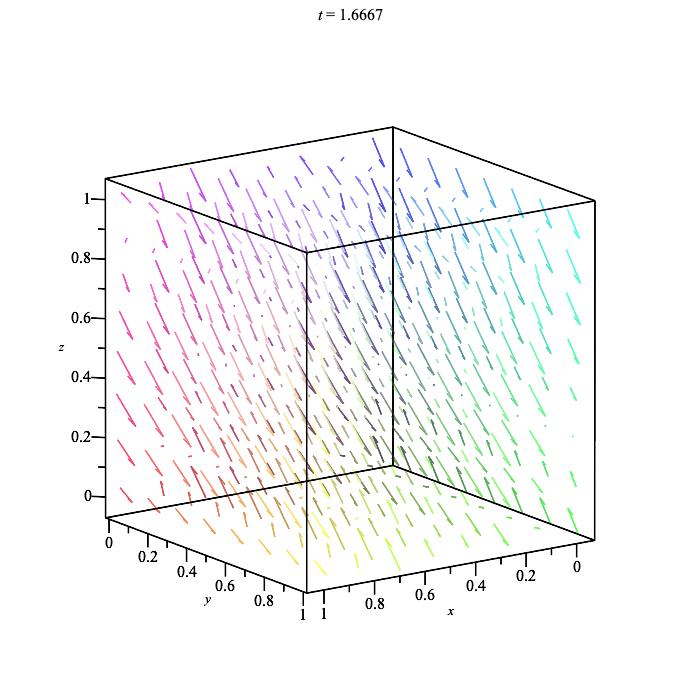}}
\subfigure{\includegraphics[scale=0.2]{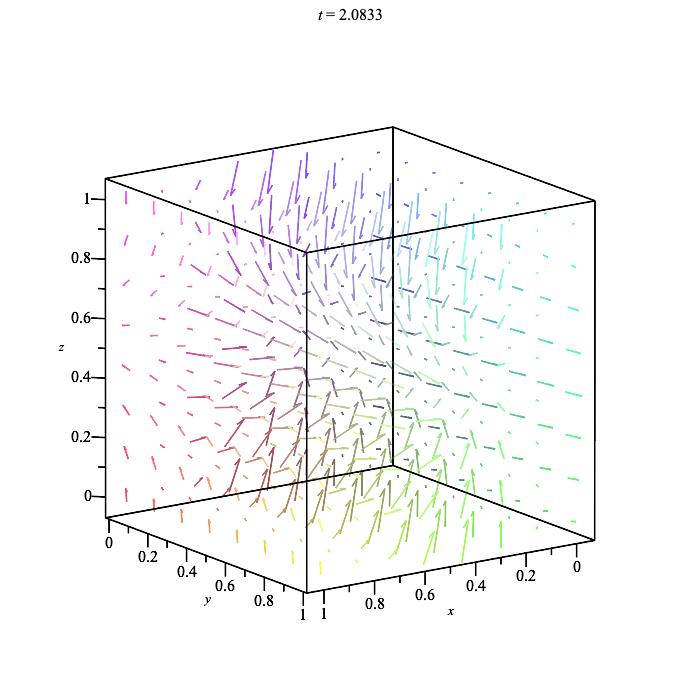}}
\subfigure{\includegraphics[scale=0.2]{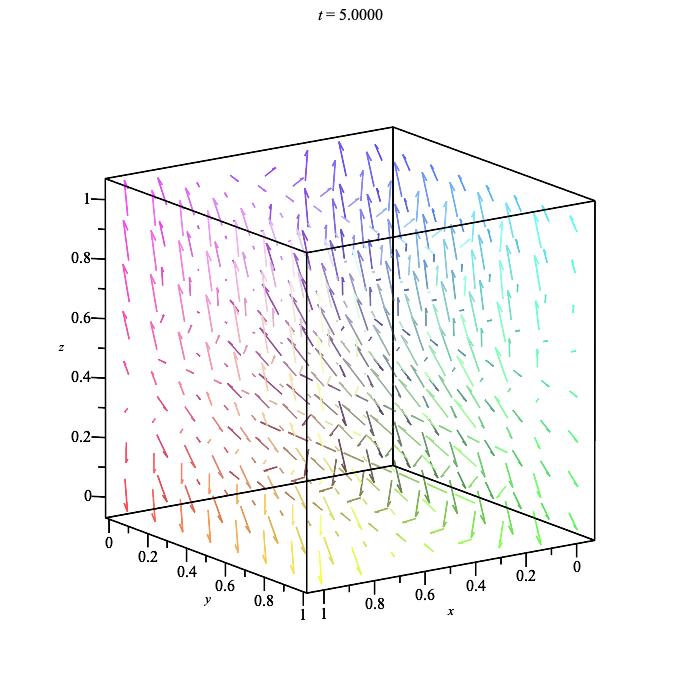}}
\subfigure{\includegraphics[scale=0.2]{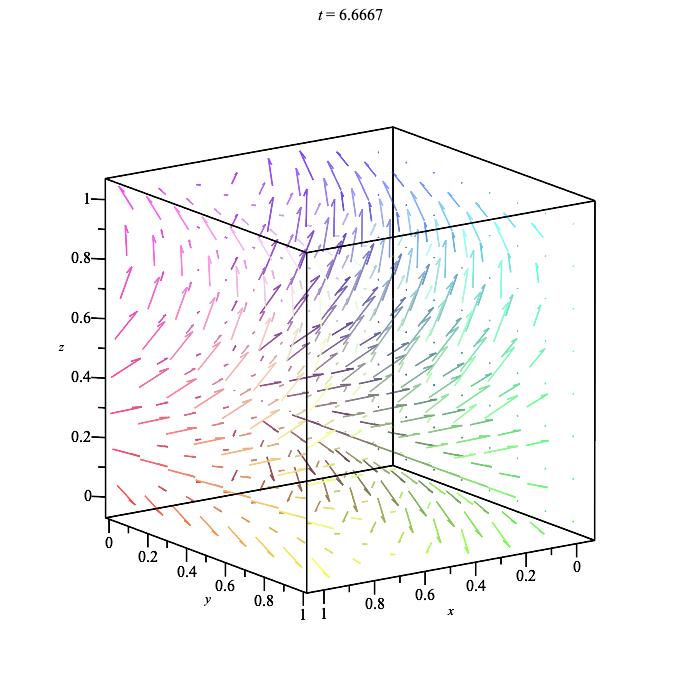}}
\subfigure{\includegraphics[scale=0.2]{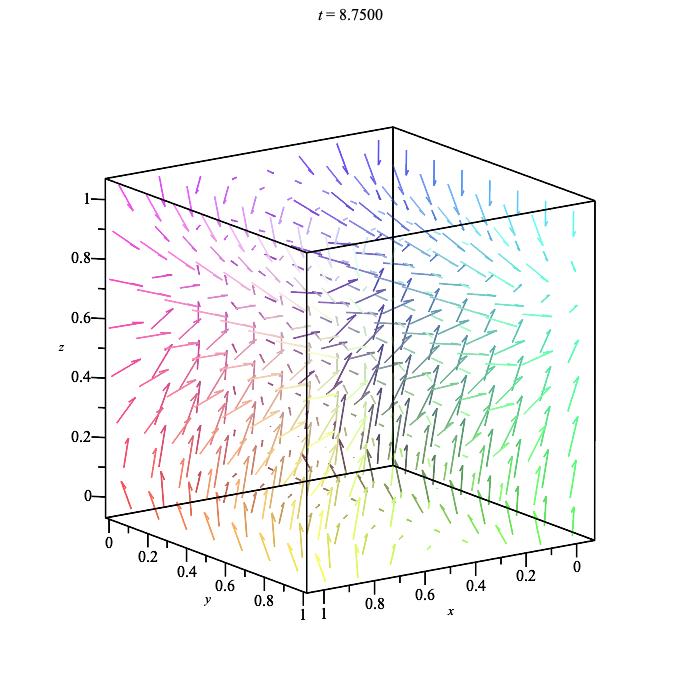}}
\subfigure{\includegraphics[scale=0.2]{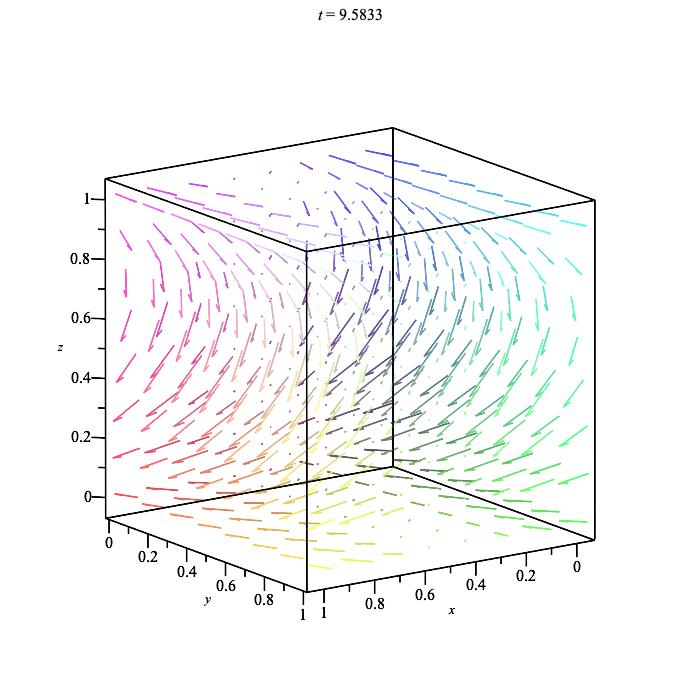}}
\caption{Plot of the Poynting vector in time}
\label{fig:poynt}
\end{figure}

To find the propagation of the perturbed EM field, we calculate the Poynting vector of the perturbation solution, as the $T^{0i}$ components of the Energy Momentum tensor.
The EM energy momentum tensor is given by
\be
T^{\mu \nu}=- \frac1{4\pi} \left(g^{ \mu \alpha} F_{\alpha \lambda} F^{\lambda \nu} + \frac14 g^{\mu \nu} F_{\alpha \beta} F^{\alpha \beta}\right).
\ee
This formula is taken from \cite{jkson} but differs from that by a minus sign as we use a metric of signature (-,+,+,+) instead of (+,-,-,-) as in that book. 
Setting specific indices, one gets the components as:
\bea
T^{01} &= & \frac{1}{4\pi} \left((1-h_{+})((1+h_+) F_{02}F_{21}+F_{31}F_{03})\right)\\
T^{02} &=& \frac{1}{4\pi} \left((1-h_+)(1+h_+) F_{12} F_{01} + (1+h_+)F_{32}F_{02}\right)\\
T^{03}&=&\frac{1}{4\pi} \left((1-h_+) F_{13} F_{01} + (1+h_+)F_{23} F_{02}\right).
\eea 
 At the linearized level, the energy momentum non-zero components are in the original x-direction, and in the z-direction. If we compute them explicitly then
{\small  \bea
 T^{0x} &= & \frac{1}{4\pi} \bigl[ B_{0y}^2 \cos^2(\omega_e x- \omega_e t)(1- A_+  \cos(\omega_g z -\omega_g t)) \bigr. \nn \\ &+& \frac{B^2_{0y} A_+ \cos(\omega_e x-\omega_e t) }{4\omega_e \omega_g}\left\{ \right.\left(\frac{\omega_g^3 - 2 \omega_e^3 -\omega_e^2\omega_g}{||\omega||} \right)\sin(\omega_e x + \omega_g z) \sin(||\omega|| t) \nn \\ &-& \left(\frac{\omega_e^2\omega_g - 2\omega_e^3 -\omega_g^3}{||\omega||}\right)\sin(\omega_e x -\omega_g z) \sin(||\omega|| t)   +\left(\omega_g^2 - 2 \omega_e^2 - \omega_g\omega_e\right) \cos(\omega_e x +\omega_g z) \cos (t ||\omega||) \nn \\ &-& \left(\omega_g^2 - 2\omega_e^2 + \omega_g \omega_e\right) \cos(\omega_e x - \omega_g z) \cos(t ||\omega||)  +  \left(\omega_g^2- 2\omega_e^2+\omega_g\omega_e\right) \cos(\omega_e x  -\omega_g z - t\ \tilde{\omega}) \nn \\ &-& \left(\omega_g^2- 2\omega_e^2-\omega_e\omega_g\right) \cos(\omega_e x +\omega_z z - t\ \bar{\omega})\left. \left. \right\}\right]
 \eea }
 Where we have used the notation $||\omega||=\sqrt{\omega_e^2+\omega_g^2}$ and $\bar{\omega}=\omega_e+\omega_g$ and $\tilde{\omega}=\omega_e-\omega_g$. 
 The other non-zero component is 
 {\small
 \bea
 T^{0z}&=& -\frac{B_{0y}^2 A_+} {16 \pi \omega_e} \cos(\omega_e x -\omega_e t) \left[ \bar{\omega} \cos(\omega_e x + \omega_g z) \cos(||\omega|| t) +\tilde{\omega} \cos(\omega_e x -\omega_g z) \cos( ||\omega|| t) \nn \right. \\  &+ & ||\omega|| \left(\sin(\omega_e x + \omega_g z) + \sin(\omega_e x-\omega_g z)\right) \sin (||\omega|| t)-\bar{\omega} \cos(\omega_e x + \omega_g z- t \bar\omega) - \tilde{\omega} \cos(\omega_e x - \omega_g z - t \tilde{\omega})\left.\right]  . 
 \eea} 
 We make a field plot of the Poynting vector perturbations in Figure (\ref{fig:poynt}) to show that there should be significant deviation from the initial direction if one is able to adjust the wave frequencies accordingly, and measure the perturbations. We again take the frequencies to be $\omega_e=4 {\rm Hz},\omega_g=3 {\rm Hz}$. The plots are of various instants of the field flow of the EM field momentum.
 
\begin{figure}  
 \includegraphics[scale=0.4]{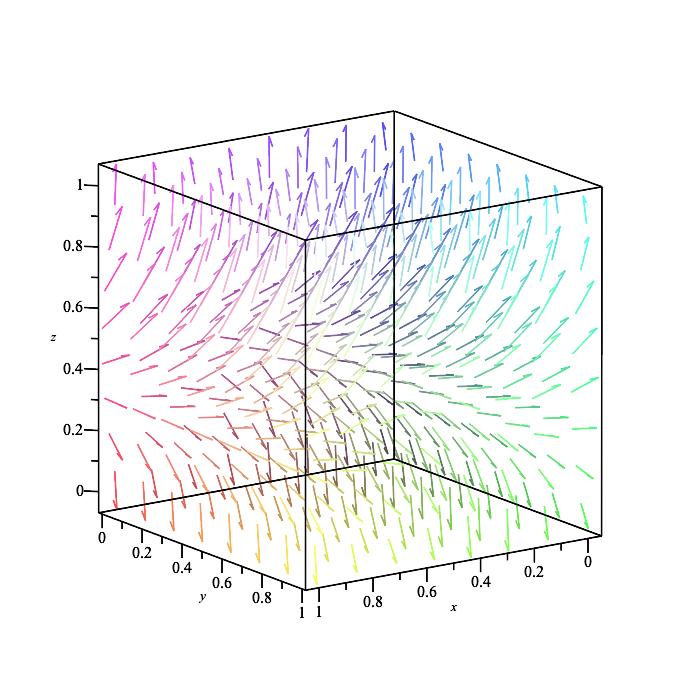}
 \caption{Time Average Poynting vector Field}
 \label{fig:poynt1}
\end{figure}

 The plot of the time snapshots as shown above Figure (\ref{fig:poynt}) of the perturbation Poynting vector shows that the field fluctuates in direction. To see if the fluctuations average to zero, we take a time integral of the momentum over the time period of the `frequency mode $\sqrt{\omega_e^2+\omega_g^2}$. The time average Poynting vector $\bar{T}^{0i}=\int_{0}^{2\pi/||\om||} T^{0i} \ dt$ plot in Figure (\ref{fig:poynt1}) shows that there is an over all bending of the EM wave at an angle which can be detected in a Maser detector \cite{maher}.
 
We further examine propagation of waves in the same direction, and find that there is no interaction. We take a magnetic field of the form $\vec{B}=B_{0y} \cos(\omega_e z -\omega_e t) \hat{y}$ and the corresponding $A_{x}$ and $E_{x}$ fields, and the equation (\ref{eqn:maxwell}) shows that there is no source for the perturbation fields.  A study of the interaction of the waves in anti-parallel directions shows that there is interaction. For this we take the magnetic field to be of the form $\vec{B}= B_{0y} \cos(\omega_e z + \omega_e t) \hat{y}$.
 
 For the example of waves travelling in the opposite direction we have this form of the perturbation:
 \bea
 \tilde{A_x} & = & \frac{A_+ B_{0y} \ \omega}2 \left[\frac{1}{\bar{\omega}+\tilde{\omega}} \left(\frac1{\bar{\omega}} \sin(\bar\omega t + \bar\omega z) + \frac1{\tilde{\omega}} \sin(\tilde\omega t + \tilde \omega z)\right) \right. \nn \\
 &+& \frac{1}{\bar{\omega} -\tilde{\omega}} \left( \frac{1}{\bar{\omega}} \sin(-\bar{\omega} t + \bar{\omega} z) - \frac{1}{\tilde{\omega}} \sin(-\tilde{\omega} t + \tilde{\omega} z) \right) \nn \\
 &-& \frac{1}{(\bar{\omega} + \tilde{\omega})} \left(\frac{1}{\bar{\omega}}\sin(\bar{\omega} z - \tilde{\omega} t) + \frac{1}{\tilde{\omega}} \sin(\tilde{\omega} z - \bar{\omega}t )\right) \nn \\
 &+& \left.\frac{1}{\bar{\omega}-\tilde{\omega}} \left(\frac{1}{\tilde{\omega}} \sin(\tilde{\omega} z - \bar{\omega} t) - \frac{1}{\bar{\omega}} \sin(\bar{\omega} z - \tilde{\omega} t)\right)\right].
 \eea
 
 where $\bar{\omega}=\omega_g+\omega_e , \tilde{\omega}=\omega_g-\omega_e$ and the $\tilde{A}_y, {\tilde A}_z$ component can also be similarly derived.
 When $\omega_g=\omega_e=\omega$ the exact solution gives us
 \be
 \tilde{A}_x = A_+ B_{0y} \frac{1}{8 \omega} \left[4( \omega t - \sin(2\om t)) + \sin(2\om (z+t)) + \sin(2\om(z-t))  - 2 \sin(2 \om z) \right]
 \ee
  There is a linear term in the perturbed mode, but we can trust the solution as far as the approximation $\tilde{A}_{x}\ll \bar{A}_x$ is true. Also as the $\vec{E}$ and ${\vec B}$ fields do not see the linear mode, the `resonance' which is characterized by runaway amplitude does not exist for the EM wave interaction with the gravitational wave.

\section{Interaction in de-Sitter background}
It is expected that gravitational waves were generated in the early universe. The remnants of the EM radiation from that time are manifest in the cosmic microwave background (CMB) which is thermalized. Primordial gravitational waves are also there in the background, but too weak to be detected yet \cite{nano}, indirect evidence of the waves are expected in the B-modes of the CMB.  However, the study of interaction of gravitational waves and matter in cosmological background is yet to be explored in details. In this paper we study the interaction of EM fields with gravitational waves in the background of conformal de-Sitter background. 
For this particular study we take the cosmological metric with a flat spatial slice, and study the system in a conformal time coordinate frame. In this frame, the metric of space-time is given by
\begin{equation}
ds^2= a^2(\tau)\left(- d\tau^2 + dx^2 + dy^2 + dz^2\right)
\end{equation}
where the scale factor is $a(\tau)$ and $\tau$ is conformal time. 
The conformal time is related to co-moving time $t$ as $-(1/H)\exp(-H t) = \tau$, where $H$ is the Hubble constant.  The scale factor is $a(\tau)= -1/H\tau$. Note that when $t\rightarrow \infty$ ; $\tau=0$  therefore conformal time is negative $-\infty<\tau<0$ as $-\infty<t<\infty$. However, for a careful analysis of the cosmology; one must include the volume factor which is a function of the scale factor, and the scale factor goes from $0$ to $\infty$ as expected from a big-bang scenario.  $a(\tau\rightarrow -\infty)\rightarrow 0$ beginning ; and expanding $a(\tau\rightarrow 0^{-})\rightarrow \infty$ as expected from the cosmological arrow of time . We begin with solving the EM waves in the cosmological metric. This gives the solutions as given they are of the form $\bar{A}_{\mu} (\tau, \vec{x}) = \bar{A}_{\mu} (\tau)\  e^{i \vec{k_e} \cdot \vec{x}} $. Where $\vec{k_e}$ is the momentum wave vector of the EM field and $k_e= \sqrt{k_{ex}^2+k_{ey}^2+k_{ez}^2}$. We use Lorenz gauge and the solutions to the Maxwell's equations in the de-Sitter background are as follows:
\bea
\bar{A}_0 (\tau)& = & \left(\frac{2}{\pi k_e}\right)^{1/2} \tau \sin(k_e \tau)  \nn \\
\bar{A}_1 (\tau)&=& P_1 \sin(k_e \tau) + Q_1 \cos(k_e \tau) - i k_{e 1} \left(\frac{2}{\pi k_e^3}\right)^{1/2} \tau \cos (k_e \tau)\nn \\   
\bar{A}_2 (\tau) &=& P_2 \sin(k_e \tau) + Q_2 \cos(k_e \tau) - i k_{e 2} \left(\frac{2}{\pi k_e^3}\right)^{1/2} \tau \cos(k_e \tau) \nn \\
\bar{A}_3 (\tau) &=& P_3 \sin(k_e \tau) + Q_3 \cos(k_e \tau) - i k_{e 3} \left(\frac{2}{\pi k_e^3} \right)^{1/2} \tau \cos(k_e \tau)
\eea
$P_i,Q_i$ are constants, fixed by the boundary conditions.
In presence of the gravitational wave the background is of the form:
\be
ds^2 = a^2(\tau)\left(-d\tau^2 + dx^a dx^b + h_{ab} dx^a dx^b\right).
\ee
The gravitational waves are of the same polarizations as in flat space-time, the $h_{+}$ and the $h_{\times}$. The exact form of the wave functions are different. If we take a wave propagating in the z-direction, then the 
\be
h_+(z,\tau)=A_+  \left({\omega_g \tau}-i\right)e^{-i \omega_g \tau} e^{i \omega_g z}
\ee
where $\omega_g$ is the frequency of the gravitational wave.
 For this paper, we discuss the effect of the $h_+$ mode on the EM fields. 
 For details of the solution for the gravitational wave in de-Sitter background see \cite{allen1,allen2}.
 
The perturbation equations for the gauge fields in the cosmological wave background with a gravitational wave are (as the Maxwell's equation are conformally invariant, one can take the equations from the previous section): 
\begin{equation}
\eta^{\alpha \mu} \eta^{\beta \nu} \partial_{\beta} F_{\mu \nu}= \eta^{\alpha \mu} h^{\beta \nu} \partial_{\beta} F_{\mu \nu} + h^{\alpha \mu} \eta^{\beta \nu} \partial_{\beta} F_{\mu \nu} + \eta^{\beta \nu} F_{\mu \nu} \partial_{\beta} h^{\alpha \mu}
\label{eqn:maxwell2}
  \end{equation}
In addition to the above one has to impose the Lorenz gauge condition for the EM gauge field $\nabla^{\mu} A_{\mu}=0$. The equation for the 0th component of the perturbation can be obtained by setting $\alpha=0$. This gives from Equation (\ref{eqn:maxwell2})
\be
\eta^{00} \eta^{\beta \nu} \partial_{\beta} \tilde{F}_{0\nu} =\eta^{00} h^{\beta \nu} \partial_{\beta} {\bar F}_{0\nu}. 
\ee
(where $\tilde F$ is the perturbation and $\bar{F}$ is the background.)
Setting $\alpha=i=1,2,3$ one gets for the LHS of Equation (\ref{eqn:maxwell2}):
\bea
\eta^{ii}\eta^{\beta \nu} \partial_{\beta} F_{i\nu} &= &\eta^{ii}\eta^{00} \partial_i \partial_0 A_0 -\eta^{ii}\left(\eta^{00} \partial_0^2 + \eta^{jj}\partial_j^2\right)A_i +\eta^{ii}\eta^{jj} \partial_i \partial_j A_j\\ 
&=& \eta^{ii}\eta^{00} \partial_i \partial_0 A_0 -\eta^{ii} (\square A_i)+ \eta^{ii}\partial_i \left(2 A_0 {\cal H} +\partial_0 A_0 -h_+ (\partial_2 A_2 -\partial_1 A_1) + h_{\times} (\partial_1 A_2 +\partial_2 A_1)\right)
\eea
In the second equation we have used the Lorenz gauge condition $\nabla_{\mu} A^{\mu}=0$. 
This gives for Equation (\ref{eqn:maxwell2}) (${\cal H}=\frac{\dot{a}}{a}$ is the Hubble's parameter)
\be
-\eta^{ii} \square A_i + \eta^{ii} 2 {\cal H} \partial_i A_0 =  \eta^{ii} \partial_i (h_+ (\partial_2 A_2 -\partial_1 A_1) )-\eta^{ii}\partial_i (h_{\times} (\partial_1 A_2 + \partial_2 A_1) )+\eta^{ii} h^{\beta \nu}\partial_{\beta} F_{i \nu} + h^{i \mu} \eta^{\beta \nu} \partial_{\beta} F_{\mu \nu} +\eta^{\beta \nu} F_{\mu \nu} \partial_{\beta}h^{i \mu}.
\ee

Using the example of the EM field where the constants $P_i,Q_i$ are zero, we get the non-zero components of the Electric field strength to be
$$\bar{F}_{0i}=-i \frac{k_{e i} }{k_e} \left(\frac{2}{\pi k_e}\right)^{1/2} \cos(k_e \tau) e^{i \vec{k_e}\cdot \vec{x}}. $$
This boundary condition of this form, produces spontaneous `EM radiation'.
Using that one gets for the perturbations in the gravitational wave background as (without the $h_{\times}$) :

 For the zeroeth component one has
 \bea
 \square \tilde{A}_0 - 2 \dot{\cal H} \tilde{A}_0 - 2 {\cal H} \partial_0 \tilde{A}_0 &= & \partial_0 h_+ (\partial_1 \bar{A}_1 -\partial_2 \bar{A}_2) + h_+ (\partial^2_1 -\partial_2^2) \bar{A}_0\\
 &=& A_+ e^{-i \omega_g \tau} (\omega_g \tau)e^{i\omega_g z} (- i \omega_g ) (k_{e 1}^2 -k_{e 2}^2 )\left(\frac{2}{\pi k_e^3}\right)^{1/2} \tau \cos (k_e \tau) e^{i \vec{k_e}\cdot\vec{x}} \nonumber \\ && - A_+  e^{-i \omega_g \tau}\left(\omega_g \tau -i\right) e^{i \omega_g z} (k_{e 1}^2 -k_{e 2}^2) \left(\frac{2}{\pi k_e}\right)^{1/2} \tau \sin(k_e \tau) e^{i \vec{k_e} \cdot \vec{x}}
 \eea
 We assume motivated from the RHS of the above equation the $\tilde{A}(\tau,\vec{x}) =\tilde{A}(\tau) \ e^{i \vec{\tilde{k}}\cdot x}$ where $\vec{\tilde{k}}=\vec{k}_g+\vec{k_e}$. Plugging this we get an equation for $\tilde{A}_0$ which is a pure function of $\tau$ of the form
 \be
 \frac{d^2{\tilde{A}_0(\tau)}}{d\tau^2} -\frac{2}{\tau} \frac{d{\tilde{A}_0(\tau)}}{d\tau} +(\tilde{k}^2+ \frac{2}{\tau^2})\tilde{A}_0 (\tau)= -g(\tau)
 \ee
 where
 \bea
 g(\tau) &= & A_+ e^{-i \omega_g \tau} \omega_g\tau(- i \omega_g ) (k_{e 1}^2 -k_{e 2}^2 )\left(\frac{2}{\pi k_e^3}\right)^{1/2} \tau \cos (k_e\tau)  \nonumber \\ && - A_+  e^{-i \omega_g \tau}\omega_g\tau \left(1-\frac{i}{\omega_g \tau}\right)  (k_{e 1}^2 -k_{e 2}^2) \left(\frac{2}{\pi k_e}\right)^{1/2} \tau \sin(k_e \tau) .
 \eea
 Using MAPLE we obtain the solution to the above equation as
 {\small \be
 \tilde{A}_0= C_0\tau\cos(\tilde{k}\tau) + D_0 \tau\sin(\tilde{k}\tau) - \frac{\tau}{\tilde{k}} \left(\sin(\tilde{k} \tau)\int\frac{\cos(\tilde{k} \tau) g(\tau)}{\tau} \ d\tau - \cos(\tilde{k}\tau)\int \frac{\sin(\tilde{k}\tau)g(\tau)}{\tau} \ d\tau\right)
 \ee}
 where $C_0, D_0$ are constants. 
 We find the re-appearance of the `new mode' frequencies of $\tilde{k}=\sqrt{k_g^2+k_e^2}$ and the integrals can be obtained using MAPLE, and the answer is obtained in rather complicated form given in the Appendix (\ref{eqn:a0}). Using MAPLE plot as a function of $\tau$ (using $C_0=D_0=0$) we find that the gravitational wave produces modulation over the shape of the unperturbed wave. Note this perturbation is proportional to $A_+$ and therefore much weaker than the original wave. We expect to study a way to make these modulations detectable, and its effect on the CMB in the near future.
 
 For the plot we use $k_{e 1}=4, k_{e 2}=0,\omega_g=3$ and the frequency of output mode as $\tilde{k}=5$. The solution from Equation (\ref{eqn:a0}) 
takes the following form:
\be
{\tilde A}_0(\tau) \sim C_0 \tau \cos(5\tau)+ D_0 \tau\sin(5\tau) - \frac{4}{\sqrt{2\pi}} A_+\tau e^{-3 i \tau}\left(\left( 2i \tau +\frac{4}{3}\right)\cos(4\tau) +\frac{( -i+3)}{2}\sin(4\tau)\right)
\ee
We have used the similarity sign as there can be some normalization constants.
We then find the mode for the propagation of the $\tilde{A}_1$ component. The Equation for the $\tilde{A}_1$ is found as

\be
\square \tilde{A}_1 = 2 {\cal H} \partial_1 \tilde{A}_0 + h_+ \partial_1^2 \bar{A}_1-h_+\partial_0 \bar{F}_{01} - \bar{F}_{01} \partial_0 h_+
\ee
Using the same method as for the $\tilde{A}_0$ we find that the time dependent part of the equation can be obtained as:
{\small \be
\partial_0^2 \tilde{A}_1(\tau)+ \tilde {k}^2 \tilde{A}_1(\tau) =-\frac{2}{\tau} (i k_{ e1}) \tilde{A}_0 (\tau)  -A_+ \frac{i k_{e 1}}{k_e} \left(\frac{2}{\pi k_e}\right)^{1/2} \left[\left( k_{e 1}^2(\omega_g \tau-i)  + i \omega_g^2\right)\tau\cos(k_e\tau) + (\omega_g \tau-i)k_e \sin(k_e\tau)\right]e^{- i \omega_g \tau}
\ee }
As previously we take the example of $k_{ e1}=4, k_{ e2}=0,\omega_g=3$ and find the following solution for $\tilde{A}_1 (\tau)$ gives
{\small \be
\tilde{A}_1(\tau) \sim C_1 \cos(5 \tau)+D_1 \sin(5 \tau) - \frac{2 A_+}{\sqrt{2\pi}} \left(\left(\frac{1}{3}(1-i) +i \tau^2+ \left(\frac{13}{16}+ \frac{i}{3}\right)\tau\right)\sin(4\tau) +\frac14\left(i \tau-\frac{25}{48} + \frac{2i}{3}\right) \cos(4 \tau)\right)\ee}
If we set $k_{e 2}=0$ then 
\be
\square{\tilde A}_2=0
\ee
if we keep $k_{e 2}$ then the equation is similar to that for $\tilde{A}_1$. Further:
{\small \bea
\square \tilde A_3 (\tau) & = & -2 {\cal H} \partial_3 \tilde{A}_0 + \partial_3 h_+ (\partial_2 \bar{A}_2 - \partial_1 \bar{A}_1)\sim -2 \frac{i \omega_g} {\tau} \tilde{A}_0 + i \omega_g e^{- i \omega_g \tau} (\omega_g \tau -i) (k_{e 2}^2 - k_{e 1}^2) \tau\cos(k_e \tau) 
 \eea}
 
 We can solve for $\tilde{A}_3$ easily using MAPLE and we get:
 \be
 \tilde{A}_3(\tau) \sim  C_3 \cos(5\tau)+ D_3 \sin(5 \tau) - \frac{2 A_+}{\sqrt{2\pi}}  e^{-3 i \tau} \left[\left(-\frac{3\tau}{4} + \frac{i}{4} +\frac{3}{8}\right)\cos(4\tau) + \left(-\frac32 \tau^2 +\frac{3}{16}  + i \tau\right) \sin(4\tau) \right]   
\ee

\begin{figure}
\subfigure{\includegraphics[scale=0.3]{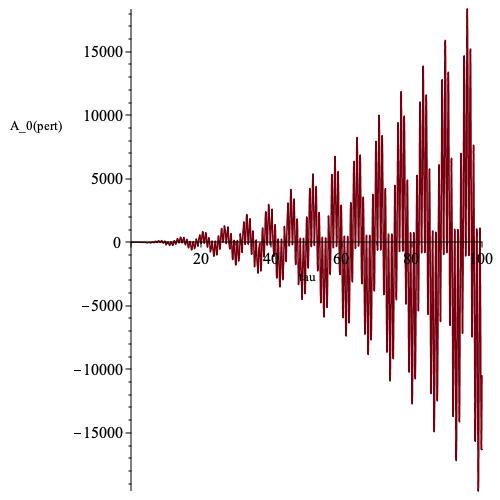}}
\subfigure{\includegraphics[scale=0.3]{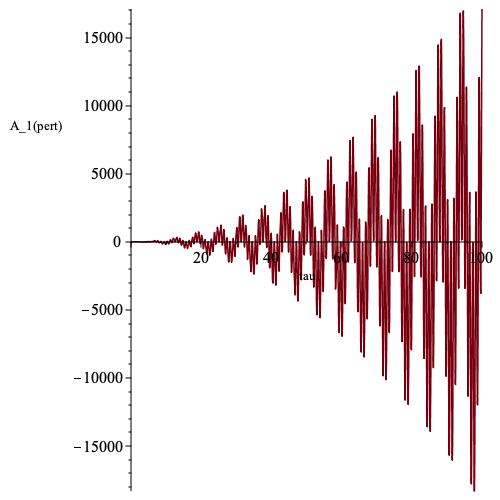}}
\subfigure{\includegraphics[scale=0.3]{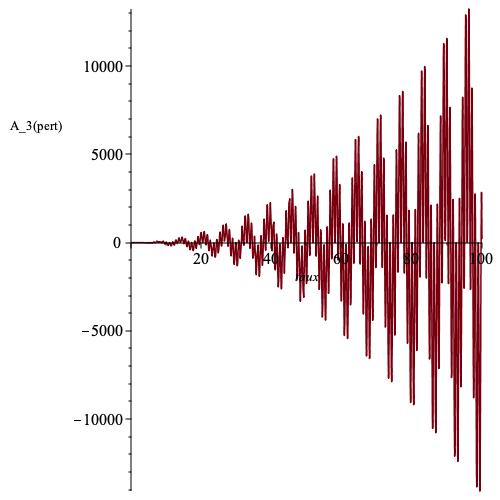}}

\caption{The perturbed Gauge field (real part) as a function of increasing conformal time, which show oscillation in time as a response to the gravitational wave.}\label{fig:cosmo}
\end{figure}
\begin{figure}
\includegraphics[scale=0.3]{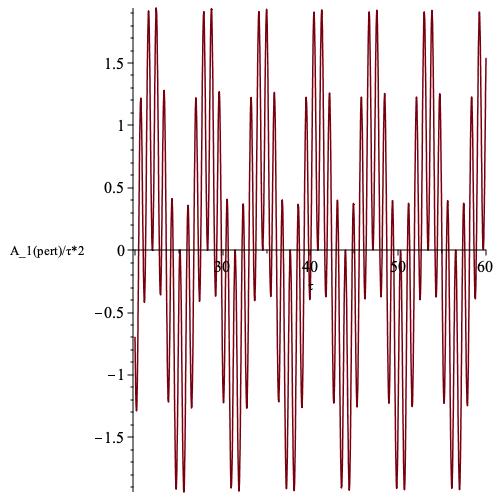}
\caption{$\tilde{A}_1/\tau^2$ which identifies the time period of the pattern as 6.27 s, The plot is for $\tau=20s..60 s$.}
\label{fig:time}
\end{figure}

As expected, the fields grow, but unlike the zeroeth order solution, the perturbed gauge potential grows as $\tau^2$. This is a significant result, and as the perturbation grows faster than the unperturbed Gauge field, the physics of the perturbation will be important very quickly, and the perturbation approximation will break down. In the above analysis however, we are simply comparing the functional form of the perturbations with the zeroeth order, where the gauge field grows linearly with $\tau$. In the plots the arrow of increasing conformal time is not the same as the increase in cosmological scale factor. The cosmological arrow of time is different, as a function of the conformal time $a(\tau)=-1/H\tau$, and the behaviour of the gauge fields will be interpreted accordingly. However, what is interesting is that there is a clear distinct pattern which is repetitive in the plots. The pattern can be identified by modding out the $\tau^2$ growth. A typical sample of the perturbation of $\tilde{A}_1$ is given in the Figure (\ref{fig:time}). The pattern period is approximately $\sim 6.27 s(\pm .02s) $ for each of the gauge fields. The above can be identified with a frequency of $\omega=2\pi/6.27 Hz\approx 1.001 Hz$. 
This is not surprising as this is the same as the difference in frequency of the EM and the gravitational waves, and this is a natural phenomena in the superposition of two waves. Here of course there is more to the story as the interaction causes the amplitude to grow. The pattern is also different from that of a simple sinusoidal wave, and this frequency represents the profile of the wave and its propagation in time. If we analyze this for realistic situation then the typical EM frequency far exceeds the Gravitational wave frequency. To find any tangible effect one has to use the radio frequency range of the CMB spectrum as in the flat space example. Further, as there are no external sources for the EM fields due to the boundary condition $P_i=Q_i=0$, the EM perturbations are generated spontaneously from the gravitational waves and the cosmological background. This phenomena when quantized will correspond to particle creation in the cosmological background. Note that the exact nature of the perturbation should be evaluated using the Green's function for the EM field in de-Sitter background as in \cite{morales}. The Green's function can be found in \cite{woodard}, however, the calculations are much more complicated than that obtained using Fourier decomposition as in the above.  In this paper we have shown that the perturbations on a EM wave induced by a gravitational wave can be calculated, however we have to study the system in a more realistic situation, with the study of observables like the Poynting vector. Further one has to study the effect of this interaction on the actual thermalization process and emergence of the CMB if the inflationary scenario is true. This is work in progress. 

\section{Conclusion}
In this paper, we studied the interaction of the EM wave with the gravitational wave and obtained some interesting observations about this in two different backgrounds (i) Flat space-time, and (ii) De-sitter space. Both of the calculations were done in the linearized approximations. In flat space-time,  we find that the EM wave deviates from its original direction when the waves have a perpendicular component, and the perturbed EM wave has a new frequency. To detect both of these results we have to re-design an experiment which is work in progress \cite{maher}. Further, in this paper, the boundary conditions as required in a realistic interferometric set up \cite{chrus} have to be discussed. However, as we observe to measure the new `frequency' in an earth experiment one has to use Masers and this is being investigated \cite{maher}. The work does however clear one aspect of the EM and gravitational wave interaction, and that is the fact that when the waves are in the same direction, there is no interaction. When the waves are in opposite directions, there is an interaction, which again is measurable in an interferometer. In the example of the EM and gravitational field interaction in de-sitter space, we solve for the Fourier mode of the perturbation. We find that the Fourier mode grows quadratically in conformal time which is different from the functional behaviour of the zeroeth order gauge field which grows linearly. The frequency of the perturbed wave is affected by the gravitational wave, and this should be observable when the frequencies are comparable. The remnants of this interactions have to be found in the CMB. We just have to find a way to access them in the current day observations. \\
{\bf Aknowledgements:} Note that most of the work appears in AP's M. Sc. thesis. We would like to thank the supervisory committee members Prof. Faraoni and Prof. Seyed-Mahmoud for their positive comments on the work. 
\appendix*
\section{Appendix1}
Here we describe the solution of a inhomogeneous wave equation as described in \cite{morales}
Given a wave equation for a unknown function $\phi(t,x)$
\be
\square \phi(t,x)=h(t,x)
\ee
with the boundary conditions $\phi(0,x)=0$ and $\frac{\partial \phi(0,x)}{\partial t}=0$ , the solution using Duhamel's principle and Kirchoff's theorem is
\be
\phi(t,x)= - \frac{1}{4 \pi} \int \frac{h(t',x')}{r} \ d^3 x'
\ee
where $r=||\vec{x}-\vec{x'}||$, and using the causality restriction $t'=t-r$ (note the speed of light c=1). 
In the derivations of the solutions as in Equation(\ref{eqn:sol1}) the integrals which have been used are   
\be
\int \frac{\sin(\omega_g z'- \omega_g t')\cos(\omega_e x' - \omega_e t')}{r}  d^3 x'
\ee
writing the exponential forms of the sines and cosines, we get integrals of the form
\be
I(x,t)= \int \exp(i (\omega_g z' - \omega_g t'))\exp(i( \omega_e x' - \omega_e t')) r dr \sin\theta d\theta d\phi
\ee
where we have taken the spherical measure. We then write $z'= r\cos\theta +z, x'= r\sin\theta\cos\phi +x , t'=t-r$ in the exponents and compute the integrals.
\bea
I (x,t) & = & e^{i \omega_g(z-t)} e^{i \omega_e(x-t)} \int_{r=0}^t \int_{\theta=0}^{\pi} \int_{\phi=0}^{2\pi}  e^{i (\omega_g+\omega_e) r + i \omega_g r \cos\theta} e^{i \omega_e r \sin\theta\cos\phi} r dr \sin\theta d\theta d\phi \nn \\
&=& 2 \pi e^{i \omega_g(z-t)} e^{i \omega_e(x-t)} \int_{r=0}^t \int_{\theta=0}^{\pi} e^{i (\omega_g+\omega_e) r + i \omega_g r \cos\theta} J_0(\omega_e r \sin\theta) r \ dr \sin\theta d\theta \nn \\
&=& e^{i \omega_g(z-t)} e^{i \omega_e(x-t)} \int_{r=0}^{r=t} e^{i (\omega_g+\omega_e) r}  \frac{\sin{||\omega|| r}}{||\omega||} dr \\
&=&  e^{i \omega_g(z-t)} e^{i \omega_e(x-t)}\frac{2 \pi}{i \omega_g \omega_e} \left[-1 + e^{i (\omega_g+\omega_e) t} \left(\cos(||\omega|| t) - i \frac{(\omega_g+\omega_e)}{||\omega||}\sin(||\omega|| t)\right)\right]
\eea
  In the above we have used Gradshteyn Rhyzik (pages 491\  \& \ 722) \cite{grad} to compute the $\phi$ and the $\theta$ integral. Computing all the terms in the integral of the exponential form, we get as the Real part of the solution Equation(\ref{eqn:sol1}). 
  
\section{Appendix2}
The explicit formulas for the Gauge fields solved in the cosmological background.
{\small
\bea
 \tilde{A}_0 &\approx& \left(\frac{2}{\pi k_e}\right)^{1/2}\frac{A_+\tau \exp(-i \omega_g \tau ) k_e }{((\tilde{k}-k_e+\omega_g)^2(\tilde{k}-k_e-\omega_g)^2(\tilde{k}+k_e+\omega_g)^2(\tilde{k}+k_e-\omega_g)^2)}  \left[k_e\left((\omega_g \tau -1) k_e^6 \right. \right. \nn \\ &+&  \left((-\tau+2)\omega_g^3  + 7i \omega_g^2 - 3{\tilde k}^2 \omega_g \tau + 3 i \tilde{k}^2\right) k_e^4 \nn \\ 
 &+ & ((-\tau -4)\omega_g^5 -3i \omega_g^4 - 2 \tilde{k}^2(\tau +2) \omega_g^3 -2 i {\tilde{k}}^2 \omega_g^2 + 3 \tilde{k}^4 \omega_g \tau - 3i \tilde{k}^4) k_e^2  \nn \\ &+&  (\tilde{k}-\omega_g)^2((\tau+2)\omega_g^3 - 3i \omega_g^2 -\tilde{k}^2 i)(\tilde{k}+\omega_g)^2\left.\right)\sin(k_e \tau) + \left(-i(2\tau+1) k_e^6 +\left((4\tau+1)i \omega_g^2 + 8 \omega_g \nn \right. \right. \\ &+& 4 i{\tilde{k}}^2 \left(\tau -\frac{3}{4}\right)k_e^4 - 2(\tilde{k}^2-\omega_g^2)\left((-i x +\frac12) \omega_g^2 - 4 \omega_g +i (\tau+\frac32){\tilde k}^2\right)k_e^2 + (\tilde{k}^2 -\omega_g^2)^3 i \left.\right)\omega_g^2 \cos(k_e \tau))\left.\right] \label{eqn:a0}
\eea}


\end{document}